\documentstyle[12pt]{article}
\def\gtwid{\mathrel{\raise.3ex\hbox{$>$\kern-.75em\lower1ex\hbox{$\sim$}}}}
\def\ltwid{\mathrel{\raise.3ex\hbox{$<$\kern-.75em\lower1ex\hbox{$\sim$}}}}

\def\cmm2{{\,\rm cm^{-2}}}
\def\cm2{{\,{\rm cm}^2}}
\def\cmm3{{\,{\rm cm}^{-3}}}
\def\gcmm3{{\,{\rm g\,cm^{-3}}}}

\def\fun#1#2{\lower3.6pt\vbox{\baselineskip0pt\lineskip.9pt
  \ialign{$\mathsurround=0pt#1\hfil##\hfil$\crcr#2\crcr\sim\crcr}}}

\begin{document}

\thispagestyle{empty}

\begin{center}

\rightline{CfPA-94-Th-39}
\rightline{astro-ph/9408094}
\rightline{Submitted to ApJ Lett. }

\vspace{0.5in}
{\Large\bf  Baryonic Dark Matter and the Diffuse $\gamma$-ray Background \\}

\vspace{.1in}
{\large\bf Xiaochun Luo  \   \ and  \   \   Joseph Silk } \\
\vspace{0.1in}

{ Center for Particle Astrophysics\\
301 Le Conte Hall\\
 University of California, Berkeley, CA  94720}\\

\end{center}

\vspace{.1in}
\centerline{\bf ABSTRACT}
\vspace{0.1in}

We show that the baryonic gas content
of the halo of our galaxy can be probed by performing a  multipole expansion
on the distribution of  diffuse background gamma-ray
emission. While the monopole moment
(isotropic background) can be used to constrain the baryonic fraction of the
halo gas, the quadrupole to monopole ratio is a sensitive probe of
the distribution of gas in the halo, {\it i.e.} the degree of flattening
of the gas distribution. The predicted diffuse gamma ray flux is found to be
very sensitive to the adopted cosmic ray density distribution throughout the
halo. If the cosmic rays are uniform, then the upper bound on the gas fraction
is  $16.6\%$ regardless of the flattening of the halo
gas distribution. However, this bound can be weakened by taking into
account the  removal of  flux in and close to the galactic plane,
 especially for a oblate
($e <1$) gas distribution. On the other hand, in the more realistic situation
that the cosmic
rays linearly  trace the smoothed halo  mass (and halo gas)
distribution, then a stringent
bound on the baryonic gas fraction in the halo $\eta$, $\eta \ltwid 3\%$, can
be placed with existing  data, regardless of halo flattening.

Subject headings: cosmology: dark matter --- Galaxy: halo --- gamma ray: theory
\newpage

\section{Introduction}
While the existence of dark matter is now firmly established, however,
despite decades of herculean efforts, its nature
remains as elusive as ever.
Indeed it is  plausible that there is more than one type
of dark matter, so that different
types of dark matter dominate on different length scales.
 In this paper, we will focus
on the smallest scale in the problem: the galactic halo scale, and
 address the following two questions about halo dark matter:
 (1) What fraction of
dark matter in the  halo is in the form of  baryonic cold gas?  and
(2) How is this gas distributed in the halo?
 We will show how   measurements of the diffuse
gamma-ray background can lead to a quantitative understanding of both
questions. Our study is motivated by recent suggestions that the
halo dark matter may consist of  dense molecular clouds
(Pfenniger, Combes and Martinet 1994; Pfenniger \& Combes 1994; Gerhard \& Silk
1994).

Several recent papers have discussed the idea of using the diffuse gamma ray
flux to
constrain the fraction of halo diffuse gas (Gilmore 1994; De Paolis et al.
1994). In this paper, we use  a detailed gamma-ray production function and two
models of cosmic
ray distribution to obtain a more accurate estimate of the diffuse
gamma ray flux from
baryonic gas. Furthermore, a   multipole expansion of the
diffuse gamma ray background is developed with the aim of  exploring the
possibility of detecting the flattening of  halo gas distribution by
measuring the multipole moments of  the gamma ray background.
\section{Gamma-ray Production Function}
The production of gamma rays through the interaction of cosmic rays with
diffuse interstellar matter and  the ambient photon field is well understood
(Bertsch et al. 1993).
At gamma ray  energies below 70 MeV, bremsstrahlung is the dominant
production process. However, as shown in Fig. 1 of Bertsch et al. (1993), the
contribution
from bremsstrahlung is comparable to that of the nucleon interaction, and thus
it is important to include this contribution. Otherwise, the gamma-ray flux
will be under-estimated by roughly a factor of two at energy threshold $E_{c}=$
 70 MeV and a factor of 1.6 at $E_{c} =$ 100 MeV.
The typical gamma-ray energy produced
through inverse Compton scattering is  $E_{\gamma} = {4 E_{e} \epsilon_{\gamma}
/ 3 m_{e}c^{2}}$, where $E_{\gamma}$ is the gamma ray energy, $E_{e}$ is the
cosmic ray
electron energy, $m_{e}$ is the electron rest mass and $\epsilon_{\gamma}$ is
the typical energy of diffuse photons in the galaxy. To produce gamma rays
above
100 MeV, very high energy electrons are required. Since the energy threshold
of gamma ray satellites  COS B (Bloemen 1989)  and SAS 2 (Thompson \& Fichtel
1982) is above 100 MeV,  it is a good
approximation to neglect inverse Compton scattering  as a source of diffuse
high energy gamma rays.

Any  spectral and spatial variations of the proton-to-electron ratio
 may be considered to be negligible, based on a study of life-time and
secondary production
(Fichtel \& Kniffen 1984). Thus,  the total integrated  gamma ray
production function can be written as
\begin{equation}
q_{\gamma} ( > E_{c}) = \int_{E_{c}}^{\infty} [q_{n} (E) + q_{e} (E)]  dE,
\end{equation}
where $q_{n}(E)$ and  $q_{e}(E)$ are the differential energy gamma-ray
production function, per atom from interstellar material in the solar neighbour
for
nuclear interaction and electron bremsstrahlung respectively. We adopt
the values given by Bertsch et al. (1993) for $q_{n}(E)$ and $q_{e}(E)$.
The integrated gamma ray production function  depends on the threshold  $E_{c}$
of the gamma ray detector.
For $E_{c} =  70 {\rm MeV}$, $100 {\rm MeV}$, $300 {\rm MeV}$,
$ 1 {\rm GeV}$, the production function is
$q_{\gamma}( > E_{c}) =  28.9, 22.6, 8.4, 2.1$ $ \times  10^{-26}
 s^{-1} $,  respectively. We will use this gamma ray production function
to calculate the distribution of the diffuse gamma ray background.
\section{Distribution of Diffuse $\gamma$-ray Emission }
The intensity of diffuse gamma ray at galactic
longitude $l$ and latitude $b$ is expressed in  general by:
\begin{equation}
j(l, b) = {1\over 4\pi} \int c_{n}(\rho, l, b) q_{\gamma}(>E_{c})
n_{H} (\rho, l, b)  d\rho
\ \ \gamma cm^{-2} s^{-1} sr^{-1} Gev^{-1}.
\end{equation}
The integration is over the line-of-sight distance along $l$ and $b$ from
the solar origin, denoted by $\rho$ and $c_{n}(\rho, l, b)$ is the cosmic ray
nucleon
intensity relative to the local intensity, and $n_{H}(\rho, l, b)
= [n_{HI}(\rho, l, b) + n_{HII}(\rho, l, b) + n_{H_{2}}(\rho, l, b)]$
is the hydrogen density.

The largest uncertainty in estimating the diffuse gamma-ray flux
is the lack of any quantitative understanding of the distribution of cosmic
rays in the galaxy and halo. Here we adopt the model of Bertsch et al. (1993)
where the density of cosmic rays is taken to be directly
 proportional to the coarse-grained
halo matter density. The normalized cosmic ray intensity function $c(\rho, l,
b)$ is given by
\begin{equation}
c(\rho, l, b) = [2 \pi r_{0}^{2} \int n_{m} dz]_{\rm local}^{-1}
\int\int\int n_{m}(r^{\prime}, l^{\prime}, b^{\prime})dz
\times e^{(-\zeta^{2}/2r_{0}^{2})} \zeta d\zeta d\psi,
\end{equation}
where $r_{0}$ is the scale length of the coupling of matter to the
cosmic rays,   $n_{m}$ is the total matter density, the subscript 'local'
refers to the solar neighbourhood, and  $\zeta$ and  $\psi$ are
the relative distance and angle between point $(\rho, l, b)$ and $(r^{\prime},
l^{\prime}, b^{\prime})$.
In using this model,  any localized  enhancements of the cosmic ray density
through
 supernova explosions and hot OB stars  (Gilmore 1994) is neglected,
an assumption which may be  justified after smoothing the diffuse background
with a beam
size larger than the typical angular size of active star-forming regions.

Two extreme cases for the cosmic ray distribution will be considered
in this paper. The first one is for
a uniform cosmic ray density,
$c(\rho, l, b) = 1$, which corresponds to a large coherence length
$r_{0}$. This is the model on which  previous constraints on the halo gas
fraction
 are based (Gilmore 1994). We will
perform  a more detailed calculation of the diffuse gamma-ray flux and discuss
 its dependence on the flattening of the halo gas distribution and
the removal of galactic signals, effects neglected in the previous work.  Note
that the baryonic gas model of Pfenniger et al. (1994) assumes the halo gas to
be in a disk, and that of Gerhard and Silk (1994) advocates an extremely ($\sim
1:10$) flattened halo.

To try to decide  how robust our results are  with respect to the cosmic ray
distribution, we also consider a second case where  the cosmic rays
trace the smoothed matter distribution linearly (De Paolis et al. 1994), i.e.,
$c(\rho, l, b) = {n_{H} (\rho, l, b)/ n_{H} (\rho = 0)}$.
  This is  the limiting case where the coherence length of the cosmic ray
distribution
is negligibly small compared with the size of the galaxy.
These two models bracket the theoretical predictions
expected for any realistic cosmic ray intensity function.
   According to Bertsch et al. (1993), the coherence length
is required to be $r_{0} \sim 2$kpc in order to fit the gamma ray emission in
the galactic plane. The coupling scale
 is much larger than the linear size adopted for
the cold molecular  hydrogen clouds in the model of  Gerhard \& Silk (1994),
but much smaller
than the size of the galaxy.  Thus, the case where the cosmic
rays linearly trace the  smoothed matter distribution  is probably  closer to
reality.
As we will show later, this leads to a more stringent constraint
on the baryonic fraction of dark halo.

\subsection{Diffuse  Hydrogen Gas in the Halo}
The halo distribution of diffuse hydrogen gas is modeled to be
 a spheroid with flattening $e$ and core radius $r_{c} = 3.5$ kpc,
\begin{equation}
n_{m}(R, z) = n_{0} (r_{c}^{2} + R^{2} + {z^{2}\over e^{2}})^{-1}.
\end{equation}
The mass of halo hydrogen gas within radius $r (r \gg r_{c})$ of the  galactic
center  is $M_{H}  = { 4 \pi m_{H} n_{0}} r \Gamma(e),$
here $m_{H}$ is the mass of a hydrogen atom and
\begin{equation}
\Gamma(e)
= \int_{0}^{1} {d x\over (1-x^{2}) + {x^{2}\over e^{2}}} =
\left\{\begin{array}{ll} {e\over \sqrt{1-e^{2}}} {\rm tan}^{-1}
{\sqrt{1-e^{2}}/e} & \mbox{if $e<1,$} \\ 1 & \mbox{if $ e =1,$} \\
{e \over \sqrt{e^{2} -1}} {\rm tanh}^{-1} {\sqrt{e^{2}-1}/e} &
\mbox{otherwise.}
\end{array}
\right.
\end{equation}
The total mass $M_{D}$ of a dark halo truncated at radius $R_{m}$
is known through the
rotation curve, $M_{D} = v_{c}^{2} R_{max}/G,$
where $v_{c}$ is the circular velocity of a halo tracer. If the rotation
curve continues to be flat up to $R_{m}\approx 50 kpc$, then the
total dark matter mass is $M_{D} = 10^{12} M_{cdot}$ for $v_{c} = 300
{\rm km/s}$. The baryonic mass fraction of the dark halo is then $\eta =
 M_{H}/M_{D} = n_{0}  {4 \pi \Gamma(e) G/ v_{c}^{2}}.$

Parametrized in terms of $\eta$ and $e$, the halo gas distribution
is given by
\begin{equation}
n_{m}(R, z) = \eta {v_{c}^{2}\over 4 \pi \Gamma(e) G} (r_{c}^{2} + R^{2} +
{z^{2}\over e^{2}})^{-1}.
\end{equation}
The expected diffuse gamma-ray flux is estimated for two limiting cases
of the cosmic ray distribution:

(i) The density of cosmic ray is uniform throughout the halo.

The diffuse gamma-ray distribution from halo clouds is
\begin{equation}
j(l, b) = {1\over 4\pi}  \eta {v_{c}^{2}\over 4 \pi \Gamma{e} G} q_{nm}(>
E_{c}) \int_{0}^{R_{max}} {1 \over ( r_{c}^{2} + R^{2} + z^{2}/e^{2})} d\rho,
\end{equation}
Inserting  numerical values,
 the diffuse flux of gamma-rays with  energy higher than 100 MeV is
$j(l, b) = 7.24  \times 10^{-5} \eta f(l, b) \ \ \gamma cm^{-2} sr^{-1} s^{-1}.
$
where the angular distribution $f(l, b)$ is given by
\begin{equation}
f (l, b) = {1\over \Gamma(e)} \int_{0}^{1} {1 \over A + B x + C x^{2}} d x,
\end{equation}
here
\begin{equation}
A = {R_{0}^{2} + r_{c}^{2} \over R_{m}^{2}}, \ \ B = - 2 {R_{0}\over R_{m}}
\cos(l) \cos(b), \  \
C = \cos^{2}(b) + \sin^{2}(b)/e^{2}.
\label{const}
\end{equation}
It is useful to expand the background gamma-ray radiation
 into multipole moments, i.e.
\begin{equation}
f(l, b) = \sum_{l}\sum_{m=-l}^{l} a_{lm} Y_{lm}(l, b),
\end{equation}
where $Y_{lm}(l, b)$ is the spherical harmonics.  The monopole term,
\begin{equation}
a_{0} = {1\over 4\pi} \int_{0}^{2\pi} \int_{0}^{\pi} \cos(b) db dl
f(l, b),
\end{equation}
is the uniform component of the sky distribution, which we may use
to constrain the baryonic fraction of the halo.
The dipole moment $D(l, b)$ has three components,
\begin{equation}
D(l, b) = D_{1} \sin(b) + D_{2} \cos(b)\sin(l) + D_{3} \sin(b) \sin(l),
\end{equation}
where $D_{i}$, $i =1,2,3$ are the dipole coefficients.
The dipole moment of the gamma ray sky is non-vanishing even if
the gas distribution is uniform because
we are located at a distance of $8.5 {\rm kpc}$ away from the
galactic center.  The quadrupole moment $Q(l, b)$ is
\begin{equation}
Q(l, b) = Q_{1}{3\sin^{2}b -1\over 2} + Q_{2}\sin2 b\cos l + Q_{3}\sin 2b\sin l
+ Q_{4}\cos^{2}b \cos 2l + Q_{5} \cos^{2}b\sin 2l,
\end{equation}
where $Q_{i}, i =1,...,5$ are quadrupole coefficients.

Due to the symmetry of the model hydrogen distribution, the
non-vanishing multipole moments are $a_{0}, D_{3}$ and $Q_{1}$,
so the  distribution function $f(l, b)$ is given by,
\begin{equation}
f(l, b) = a_{0} + D \cos(b)\cos(l) + Q {3 \sin^{2}(b) -1 \over 2},
\end{equation}
The uniform diffuse gamma ray background $j_{0} (\gamma)$ is found
to be $j_{0}(\gamma) = 7.24 \times 10^{-5} \eta a_{0}.$
The isotropic gamma-ray emission  determined from SAS 2
is $1.2 \times 10^{-5} \gamma(E>100 Mev)$ ${\rm cm}^{-2} {\rm s}^{-1}
\rm{sr}^{-1}$. Provide that  all of this background is produced by the hydrogen
gas
in the halo, the upper limit on the halo baryonic fraction is given by
$
\eta \ltwid 0.166/a_{0}(e).$
As we show in the left panels of Fig.1, $a_{0}$ is  not a sensitive function
of the
flattening parameter $e$, which reflects the fact that the  gamma ray
flux  only traces the total column density along the line of sight.
In the flattening range $e =0.3 -1.4$, the monopole term deviates very little
from unity. This translates to a limit on the
baryonic fraction of halo as $\eta \ltwid 16.6\%$, in agreement
with Gilmore (1994).
The ratios of dipole moment $D$ and quadrupole moment $Q$ to the monopole
term are shown in the left panels of Fig.1 as a function of $e$. As expected,
the dipole
moment  is insensitive to the change of  $e$ since it reflects only the fact
that our location is off-center. On the contrary, the quadrupole moment
changes rapidly as the distribution of gas deviates from sphericity.

In practice, to estimate all of the multipole terms, the flux from
the galactic plane has to be removed very carefully. Systematic
galactic cuts need to be performed which will however inevitably remove
gamma ray flux produced by halo gas. To see how  this
affects the predictions of various multipole moments,
in the right panels of Fig. 1,
 we plot the monopole, dipole/monopole, quadrupole/monopole
after a 30 degree cut, as a function of flattening $e$.
The reduction in predicted gamma-ray flux is considerable, especially
for $e<1$. For example, after the 30 degree cut, the isotropic
gamma ray flux for $e =0.3$ is $1.4 \eta \times 10^{-5} \gamma cm^{-2}
s^{-1} sr^{-1}$, which is consistent with the SAS 2 observation
even for a gas-dominated halo.  No upper bound on halo baryonic gas
fraction
can therefore be achieved through gamma ray  measurements.

On the contrary, the quadrupole to monopole ratio is enhanced after
performing the galactic cut. This makes it more appealing to use the
quadrupole to monopole ratio to first constrain the flattening $e$ of
the halo gas distribution . Only after the  parameter $e$ is better
constrained could one hope to use the isotropic background at high galactic
latitude  to constrain the baryonic gas fraction.

(ii) Model 2: cosmic rays linearly trace the smoothed matter distribution.

The interrelation of cosmic rays with the  matter distribution
is an unsolved issue  which still lacks clear
understanding. However, it is evident that, after smoothing
the matter density over some length scale, there is a correlation with
cosmic ray density. One of the first  such  observations,
by Puget et al. (1976), found that the cosmic ray intensity has to be 1.9 to
4.8
times the local (solar system) value in a region 5 kpc away from the
galactic center. A more detailed discussion of this issue can be found
in Bertsch et al. (1993). If cosmic rays linearly trace the  mass,
the diffuse gamma-ray distribution from halo clouds is
\begin{equation}
j(l, b) = {1\over 4\pi} n_{0} q_{nm}(> E_{c}) \int_{0}^{R_{m}} { R_{0}^{2} +
r_{c}^{2} \over ( r_{c}^{2} + R^{2} + z^{2}/e^{2})^{2}} d\rho,
\end{equation}
Inserting  the numerical values,  the diffuse flux of  gamma-rays with energy
higher than 100 MeV is $
j(l, b) = 7.24  \times 10^{-5} \eta  g(l, b) \ \ \gamma cm^{-2} sr^{-1}
s^{-1},$
where the angular distribution $g(l, b)$ is given by
\begin{equation}
g(l, b) = {r_{c}^{2} + R_{0}^{2} \over R_{m}^{2}} \int_{0}^{1}
{1\over (  A + B x + c x^{2})^{2}} dx,
\end{equation}
and $ A, B, C$ are given in Eq.(\ref{const}).
Similar to what has been done in model (1), a  multipole expansion of this
 sky distribution is performed up to   quadrupole terms.
The monopole term is found to be enhanced over that in model (1) by a factor
of 20 over the range of flattening $e$ we consider. In this
model, the upper bound on the halo density is rather
stringent: $\eta \ltwid 1\%$. Less than one percent of the
dark matter could be in the form of diffuse gas which is
capable of producing gamma ray. Although the galactic cut will also reduce
the estimated monopole flux, the upper limit still remains
strong: $\eta \ltwid 3 \%$.

Another clear feature of the diffuse background gamma ray flux is the large
dipole
moment which points toward the galactic center if cosmic rays indeed  trace
the mass
distribution. The dipole drops dramatically after a 30 degree galactic cut
since it lies mainly in the galactic plane. The quadrupole term
remains large but flips sign to point towards the galactic poles.

In summary, we have  shown that much can be learnt about baryonic gas
in the halo  through a  multipole expansion
of the diffuse background gamma-ray intensity.  While the monopole moment
(isotropic background) can be used to constrain the baryonic fraction of the
halo gas, the quadrupole to monopole ratio is a sensitive probe of
the distribution of gas in the halo, i.e., the flattening parameter
of the gas distribution. We also have found that our predictions are
very sensitive to the cosmic ray density distribution throughout the
halo. If the cosmic rays are uniform, then an upper bound on the gas fraction
is found to be $16.6\%$ regardless of the flattening of the halo
gas distribution. However, this bound can be evaded by taking into
account the  removal of  flux in and close to the galactic plane,
 especially for a oblate
($e <1$) gas distribution. On the other hand, if the cosmic
rays linearly  trace the  smoothed mass distribution, then a rather stringent
bound on the baryonic gas fraction of the halo $\eta$, $\eta \ltwid 3\%$ is
placed through current data.

This work is supported by  a grant from the NSF.
\newpage

\def\ref{\par\noindent\hangindent=2pc \hangafter=1 }
\centerline{REFERENCES}

\ref
Bertsch, D.L., Dame, T.M., Fichtel, C.E., Hunter, S.D.,  Sreekumar, P.,
Stacy, J.G., \& Thaddeus, P. 1993, ApJ, 416, 587

\ref
Bloemen, H. 1989, ARAA, 27, 469

\ref
De Paolis, F., Ingrosso, G., Jetzer, P. \& Roncadelli, M. 1994, ZU-TH-18-94

\ref
Fichtel, C.E. \& Kniffen, D.A. 1984, A\&A, 134, 13

\ref
Gerhard, O.E., \& Silk, J 1994, Nature, submitted

\ref
Gilmore, G. 1994, ApJ Suppl, 92, 539.

\ref
Pfenniger, D., \& Combes, F. 1994, A\&A, 285, 94

\ref
Pfenniger, D., Combes, F. \& Martinet, L. 1994, A\&A, 285, 79

\ref
Puget, J.L., Ryter, C., Serra, G., \& Bignami, G. 1976, 50, 247

\ref
Thompson, D.J., \& Fichtel, C.E. 1982, A\&A, 109, 352

\vspace {0.4 in}

\centerline{Figure Caption}

Fig.1: The monopole, dipole and quadrupole moment of the diffuse gamma ray
background for a spheroidal halo gas distribution as a function
of the axis ratio $e$.  The isotropic background flux   is equal to the
monopole
term $a_{0}$ times a numerical constant, $7.24 \times 10^{-5} \eta$
$\gamma$ ${\rm cm}^{-2}$ ${\rm s}^{-1}$  ${\rm sr}^{-1}$, where $\eta$ is the
gas fraction of the halo. The left panels are for full sky coverage. The right
panels are for a 30 degree galactic cut.

\end{document}